\documentclass[twoside,a4paper,11pt]{sea9}
\usepackage{graphicx}
\usepackage{hyperref}
\usepackage{amsmath,amssymb,amsfonts,latexsym,cancel}
\begin{document}
\pagenumbering{arabic}
\pagestyle{myheadings}
\thispagestyle{empty}
{\flushleft\includegraphics[width=\textwidth,bb=58 650 590 680]{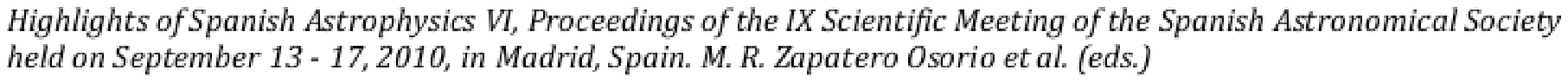}}
\vspace*{0.2cm}
\begin{flushleft}
{\bf {\LARGE
The distribution of star-forming regions in M33
}\\
\vspace*{1cm}
N\'estor S\'anchez$^{1}$,
Neyda A\~nez$^{2}$,
Emilio J. Alfaro$^{1}$,
and
Mary Crone Odekon$^{3}$
}\\
\vspace*{0.5cm}
$^{1}$
Instituto de Astrof\'{\i}sica de Andaluc\'{\i}a, CSIC, Granada, Spain\\
$^{2}$
Departamento de F\'{\i}sica, Universidad del Zulia, Maracaibo, Venezuela\\
$^{3}$
Department of Physics, Skidmore College, Saratoga Springs, USA
\end{flushleft}
\markboth{
Distribution of star-forming regions in M33
}{
S\'anchez et al.
}
\thispagestyle{empty}
\vspace*{0.4cm}
\begin{minipage}[l]{0.09\textwidth}
\ 
\end{minipage}
\begin{minipage}[r]{0.9\textwidth}
\vspace{1cm}
\section*{Abstract}{\small
We use fractal analysis to systematically study the clustering
strength of the distribution of stars, HII regions, molecular
gas, and individual giant molecular clouds in M33 over a wide
range of spatial scales. We find a clear transition from a
scale-free behavior at small spatial scales to a nearly uniform
distribution at large scales. The transition region lies in the
range $\sim 500-1000$ pc and it separates the regime of
small-scale turbulent motion from that of large-scale
galactic dynamics. The three-dimensional fractal dimension
of bright young stars and molecular gas at small spatial
scales is $D_{f,{\rm 3D}} \lesssim 1.9$ indicating that
the interstellar medium in M33 is on average much more
fragmented and irregular than the in the Milky Way.
\normalsize}
\end{minipage}

\section{Introduction}

Interstellar Medium (ISM) in the Milky Way shows fractal
patterns, that is, it is organized into irregular structures
in a hierarchical and approximately self-similar manner in which
each structure (cloud) is composed of smaller similar structures
which are composed of even smaller structures and so on. This
fractal structure is observed over a wide range of spatial scales
from $\sim 0.1$ pc to $\sim 100$ pc or even more, and it is
supposed to be a consequence of turbulent processes ocurring
in the ISM \cite{Elm04}. The formation of stars also exhibits
a spatial hierarchy ranging from the scale of a few pc for star
clusters and associations up to about a kpc for so-called star
complexes \cite{Efr95}. Fractal analysis is an appropiate tool
for characterizing these hierarchical and self-similar systems.
The fractal dimension $D_f$ quantifies the degree of irregularity
or clumpiness (spatial heterogeneity) of the distribution of gas or
stars. The more irregular or far from homogeneity is the structure,
the smaller fractal dimension values. It is often accepted that the
fractal dimension of the ISM in our Galaxy has a nearly universal
value \cite{Ber07}. From a detailed analysis of several emission
maps of three different molecular clouds, S\'anchez et al.
\cite{San07b} obtained $D_f \simeq 2.7 \pm 0.1$ with no
evidence of significant variations.

An important issue is the spatial extent of this self-similar behavior.
In the solar neighborhood, fractal behavior has been observed for the
distribution of young open cluster and young stars at spatial scales
of up to $\sim 1$ kpc \cite{Fue09}. In external galaxies, hierarchical
structures extend up to $\gtrsim 1$ kpc scales for the gas and for
stars and star-forming sites \cite{Bas09,Bon10}. However,
there seem to be variations in the fractal properties among
galaxies \cite{San08}. The distribution of gas \cite{Dut09}, stars
\cite{Ode06} and HII regions \cite{San08} seems to be
less clustered in bright galaxies. In other words, a larger star
formation rate in a galaxy tends to be correlated with a larger
fractal dimension. A significant challenge in interpreting
this and other results is that authors present measurements
for different ranges of scales and identify their samples in
different ways.
Our approach in this work is to consider a case study in which we
systematically analyze the clustering of different components
of a single galaxy over a wide range of spatial scales. Because
of its proximity, large size, and low inclination, M33 is a
suitable object for this task. Thus, here we study the
clustering strength in the distribution of young stars,
HII regions, molecular gas, and individual giant molecular
clouds (GMCs) in M33.

\section{Data and method}

\begin{figure}[ht]
\center
\includegraphics[scale=1.0]{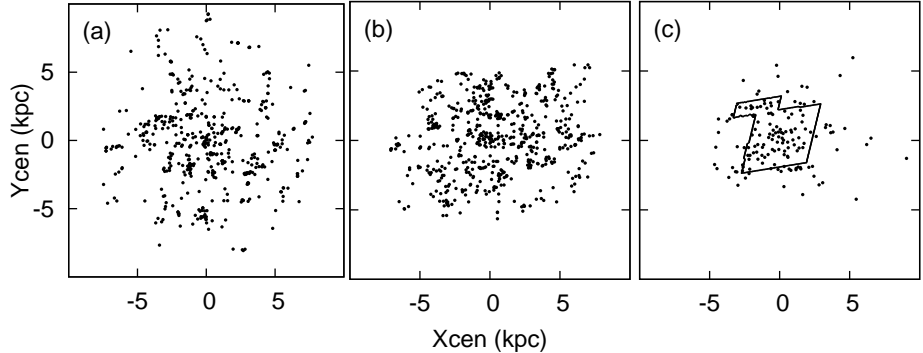}
\caption{\label{distribuciones} Spatial distributions
of (a) bright stars, (b) bright HII regions and (c) GMCs
in M33. The axis coordinates are positions relative to the
galactic center in kpc. The inset in panel (c) shows
the area corresponding to the molecular gas map.}
\end{figure}

For the distribution of stars we used the catalog of
Massey et al. \cite{Mas06} to get positions and
photometry of stars in M33. Stars were divided into two
sets that we refer to simply as ``bright" stars ($-0.3
\leq V-I \leq -0.1$ and $-6.0 \leq M_I \leq -5.0$) and
``faint" stars ($-0.3 \leq V-I \leq 0.0$ and $-4.5 \leq
M_I \leq -4.0$). The total numbers of stars for
the bright and faint sets are $534$ and $1644$,
respectively. 
For the HII regions we used the catalog of Hodge
et al. \cite{Hod99}, from which we removed regions
classified as unresolved, diffuse, linear and/or
any other factor that may raise doubts on the real nature
of such regions. We also removed the regions having null
integrated H$\alpha$ fluxes in the catalog. The total
number of ``bright" HII regions in M33 was $617$.
The distribution of $149$ molecular clouds was
obtained from the catalog of Rosolowsky et al.
\cite{Ros07}. For the molecular gas we used
high resolution CO emission data of the center region
of M33 provided by Erik Rosolowsky \cite{Ros07}. The
data cube was collapsed to produce a map of integrated
intensity with a final resolution of $20''$ ($93$ pc).
We adopted a position angle of $23$ degrees and an inclination
of $55$ degrees to deproject the positions of stars, HII regions
and GMCs. To convert angular sizes into linear sizes, we assume
a distance of $960$ kpc \cite{U09}.
The positions of bright stars, HII regions and GMCs
relative to the galactic center, and the area
corresponding to the used molecular gas map are shown
in Figure~\ref{distribuciones}.

The degree of clustering was measured by applying
algorithms that we have previously developed and
tested on simulated fractals.
For the distribution of stars, HII regions and GMCs
we calculated the correlation dimension $D_c$
from the relation $C(r) \sim r^{D_c}$, where the
correlation integral $C(r)$ is the average number
of points within a distance $r$ \cite{San08,San07a}.
In order to consider the possibility of a transition
in $D_c$ \cite{Ode08}, we performed two separate
calculations at different scales. We varied the
range of spatial scales until obtaining the best
result (the one with the minimum transition region
that minimizes the sum of the squared residuals).
For the emission map we calculated the perimeter-based
dimension $D_{\rm per}$ from the equation $P \sim
A^{D_{\rm per}/2}$ relating the perimeters $P$ and
areas $A$ of the clouds in the map \cite{San05}.

\section{Results and discussion}

\begin{figure}[ht]
\center
\includegraphics[scale=1.0]{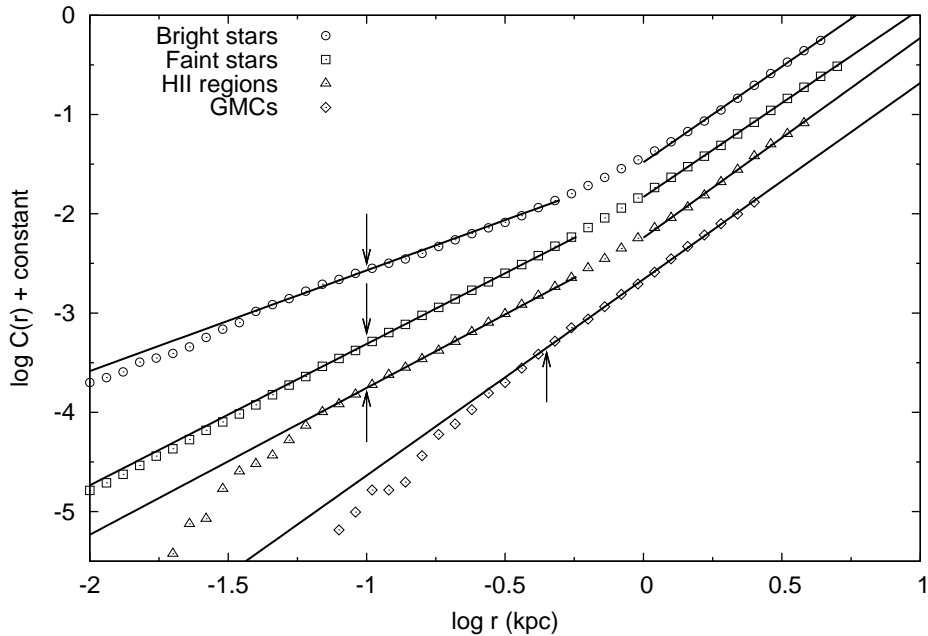}
\caption{\label{correlacion} Correlation integral
$C(r)$ for the samples of bright stars (circles),
faint stars (squares), HII regions (triangles) and
GMCs (rhombuses). The data have been arbitrarily
shifted downward (except the top one) for clarity.
Vertical arrows indicate the points above which the
algorithm gives reliable values of $C(r)$ and performs
the linear fits (solid lines). Different fits were done
below and above the range without any straight line,
except for the GMCs that do not show any change in the
slope.}
\end{figure}

Figure~\ref{correlacion} shows the correlation integral
for stars, HII regions and GMCs.
The slopes of the linear fits are the correlation dimensions
which are shown in Table~\ref{tabla}. This Table also shows
the perimeter dimension resulting from the CO emission map.
The calculated two-dimensional fractal dimensions ($D_{f,\rm 2D}$)
were converted to three-dimensional dimensions ($D_{f,\rm 3D}$)
using results from previous studies \cite{San05,San08}.
Table~\ref{tabla} shows the range of $D_{f,\rm 3D}$ values
that are compatible with the calculated values of $D_{f,\rm 2D}$.
Several interesting conclusions can be drawn from these
results, which we discuss now.
\begin{table}[ht] 
\caption{Calculated fractal dimensions for M33}
\center
\begin{minipage}{0.7\textwidth}
\center
\begin{tabular}{lcccc} 
\hline\hline 
& \multicolumn{2}{c}{Small spatial scales\footnote{For
stars and HII regions small spatial scale means $\lesssim 500$ pc
and large scale means $\gtrsim 1$ kpc. For molecular gas large scale
is $\gtrsim 500$ pc (distribution of clouds) and small scale is
$\lesssim 500$ pc (CO map).}} &
\multicolumn{2}{c}{Large spatial scales} \\
[0.5ex]
Sample & $D_{f,{\rm 2D}}$\footnote{$D_{f,{\rm 2D}}$ refers either
to the two-dimensional correlation dimension $D_c$ (for the
distribution of stars, HII regions and GMCs) or to the
perimeter-area based dimension $D_{\rm per}$ (for the CO map).
$D_{f,{\rm 3D}}$ is the corresponding three-dimensional
fractal dimension.}
& $D_{f,{\rm 3D}}$ & $D_{f,{\rm 2D}}$ & $D_{f,{\rm 3D}}$ \\
[0.5ex]
\hline 
Bright stars    & $1.01\pm0.05$ & 1.0-1.9 & $1.93\pm0.03$ & 2.8-2.9 \\
Faint stars     & $1.42\pm0.04$ & 2.2-2.4 & $1.89\pm0.02$ & 2.8-2.9 \\
HII regions     & $1.48\pm0.08$ & 2.3-2.5 & $2.01\pm0.03$ & 2.9-3.0 \\
Molecular clouds& .......       & ....... & $1.98\pm0.04$ & 2.8-3.0 \\
CO emission map & $1.65\pm0.06$ & 1.6-1.8 & ....... & ....... \\
\hline
\end{tabular} 
\end{minipage}
\label{tabla} 
\end{table}

\subsection{Transition region}

All of the objects except the GMCs exhibit scale-free
clustering on small scales and a clear transition to
a higher slope at larger spatial scales
(Fig.~\ref{correlacion}). At small scales the
correlation dimension of the distribution of stars
and HII regions is $\lesssim 1.5$ whereas at large
scales it is $\gtrsim 1.9$, and the differences are
always larger than the associated uncertainties
(Table~\ref{tabla}). The spatial scale where this
transition takes place is roughly the same for each
component ($\sim 500-1000$ pc). 
A transition from a smaller correlation dimension to
a larger one was reported by Odekon \cite{Ode08} for
young stars in M33. However, Bastian et al. \cite{Bas07}
did not find any characteristic size for the distribution of
star-forming regions. Here we provide a detailed quantification
of this transition and observe it for the first time for the
distribution of HII regions.
The transition is not observed for the distribution of GMCs
but, given the limited number of data points for this sample
($N=149$), the lower limit of reliable values is higher than
for the other objects ($r \gtrsim 500$ pc).

What is the nature of this transition? Padoan et al.
\cite{Pad01} argued that there must be a physical transition in
the statistical properties of the flow close to the disk scale
height. In turbulent flows the energy is injected at certain
spatial scale and then it ``cascades" to smaller scales. But
there are many possible energy sources that may be relevant
at different levels. A possible
consequence of this may be different distribution patterns at
different size ranges. Even though the underlying turbulent
structure tends to be the same, non-turbulent motions acting
on galactic scales could modify the final structure at those
scales. In other words, the power law behavior at small spatial
scales would be a direct consequence of the self-similar turbulent
motions in the medium, but this turbulence is unlikely to extend to
very large scales, where two-dimensional flows should dominate the
dynamics. Thus, we identify a characteristic spatial scale (around
$500-1000$ pc) that separates the regime where coherent
star formation is occurring in a turbulent medium from
the regime that is organized by large-scale galactic dynamics.
Interestingly,
the behavior we observe is that all the fractal dimensions
at $r \gtrsim 1$ kpc are within a narrow range of values
($D_{f,{\rm 2D}} = 1.9-2.0$, or $D_{f,{\rm 3D}} = 2.8-3.0$)
that are consistent with essentially uniform (random)
distributions.

\subsection{Evolutionary effects}

Young, newborn stars should reflect the same conditions of
the ISM from which they were formed. Therefore, it is
reasonable to assume that the fractal dimension of the
distribution of new-born stars should be nearly the same
as that of the molecular gas from which they are formed.
Our results are roughly consistent with this
idea within the rather large uncertainties for
$D_{f,{\rm 3D}}$. Both bright stars and molecular gas are
distributed with $D_{f,{\rm 3D}} \lesssim 1.9$. However,
faint stars and HII regions have significantly higher
fractal dimensions. We interpret these higher dimensions
in terms of evolutionary effects. It seems
that the initial clumpy distribution of
star-forming sites may evolve towards a smoother
distribution. This effect has been observed for
the distributions of young stars in LMC \cite{Bas09}
and SMC \cite{Gie08} and for the stars clusters in both
galaxies \cite{Bon10}.
It has been shown that the brightest HII regions in
spiral galaxies (which reflect, in a first approximation,
the initial distribution of star-forming sites) tend to be
distributed in more clumpy patterns than the low-brightness
regions \cite{San08}. A smoother distribution means a
higher fractal dimension, and this is what we found
for faint stars and HII regions at small scales for
which $D_{f,{\rm 3D}} \simeq 2.2-2.5$.

\subsection{Fractal structure in M33}

The three-dimensional fractal dimension of the distribution
of molecular gas in M33 is $D_{f,{\rm 3D}} \simeq 1.6-1.8$.
It is a very interesting to note that this
value is much smaller than the fractal dimension
of molecular clouds in the Milky Way, which is in the range
$D_{f,{\rm 3D}} \simeq 2.6-2.8$ \cite{San05,San07b}. That
is, molecular clouds in M33 exhibit a much more fragmented
and irregular structure than in the Milky Way.
This result may yield some clues about the main physical
processes that determine the structure of the ISM. In
principle, if the main physical mechanisms acting were
not the same we would expect different global properties
for the ISM.
Simulations of turbulent fluids produce very different structures
depending on which processes are considered in the system. For
example, Federrath et al. \cite{Fed09} showed that simulations
of supersonic isothermal turbulence in the extreme case of purely
compressive energy injection, produce a significantly smaller fractal
dimension for the density distribution ($D_f \sim 2.3$) than in the
case of purely solenoidal forcing ($D_f \sim 2.6$). Although it is
widely accepted that turbulence is the primary driver of the structure
and motion of the ISM, the main energy sources for this turbulence
are not yet well established. 
Obviously, more detailed studies are needed to clarify
this point but in this work we have found a significant
and remarkable difference in the internal structure of GMCs
between M33 and the Milky Way.

\small
\section*{Acknowledgments}
We acknowledge financial support from MICINN of
Spain through grant AYA2010-17631 and from Junta de
Andaluc\'{\i}a through TIC-101 and TIC-4075.
N.S. is supported by a JAE-Doc (CSIC) contract.
E.J.A. acknowledges financial support from the
Spanish MICINN under the Consolider-Ingenio 2010
Program grant CSD2006-00070: ``First Science with
the GTC".

\end{document}